\documentclass[12pt,preprint]{aastex}

\usepackage{amsmath}
\usepackage{natbib}

\newcommand{\kms}{km s$^{-1}$}
\newcommand{\msun}{M$_{\odot}$}

\shorttitle{ALFALFA Correlation Function}
\shortauthors{Martin et al.}

\begin{document}

\title{The Clustering Characteristics of HI-Selected Galaxies from the 40\% ALFALFA Survey}

\author {Ann M. Martin\altaffilmark{1,2},
Riccardo Giovanelli\altaffilmark{2}, Martha P. Haynes\altaffilmark{2}, Luigi Guzzo\altaffilmark{3}}
\altaffiltext{1}{NASA Postdoctoral Program, NASA Langley Research Center, Hampton VA 23618. {\textit{e-mail:}} ann.m.martin@nasa.gov}
\altaffiltext{2}{Center for Radiophysics and Space Research, Space Sciences Building,
Cornell University, Ithaca, NY 14853. {\textit{e-mail:}} riccardo@astro.cornell.edu, haynes@astro.cornell.edu}
\altaffiltext{3}{INAF Osservatorio Astronomico di Brera, Milan, Italy. {\textit{e-mail:}} luigi.guzzo@brera.inaf.it}

\begin{abstract}
The 40$\%$ Arecibo Legacy Fast ALFA (ALFALFA) survey catalog ($\alpha.40$) of $\sim$10,150 HI-selected galaxies is used to analyze the clustering properties of gas-rich galaxies. By employing the Landy-Szalay estimator and a full covariance analysis for the two-point galaxy-galaxy correlation function, we obtain the real-space correlation function and model it as a power law, $\xi$(r)=(r/r$_0$)$^{-\gamma}$, on scales $<$ 10 h$^{-1}$ Mpc. As the largest sample of blindly HI-selected galaxies to date, $\alpha.40$ provides detailed understanding of the clustering of this population. We find $\gamma$ = 1.51 $\pm$ 0.09 and r$_0$ = 3.3 + 0.3, -0.2 h$^{-1}$ Mpc, reinforcing the understanding that gas-rich galaxies represent the most weakly clustered galaxy population known; we also observe a departure from a pure power law shape at intermediate scales, as predicted in $\Lambda$CDM halo occupation distribution models. Furthermore, we measure the bias parameter for the $\alpha.40$ galaxy sample and find that HI galaxies are severely antibiased on small scales, but only weakly antibiased on large scales. The robust measurement of the correlation function for gas-rich galaxies obtained via the $\alpha.40$ sample constrains models of the distribution of HI in simulated galaxies, and will be employed to better understand the role of gas in environmentally-dependent galaxy evolution.
\end{abstract}

\section{Introduction \label{ch6intro}}

Galaxies selected by their neutral hydrogen are known to be less clustered than their optically-selected counterparts (\citet{2007MNRAS.378..301B, 2007ApJ...654..702M} for HIPASS) and less likely to be found in such dense environments. Given anticipated cosmological uses of 21 cm galaxy redshift surveys, it is important to understand the clustering characteristics of this population of galaxies. Specifically, 21 cm line surveys obtain detections and redshift concurrently, along with HI mass, reducing their expense and eliminating the need for follow-up observations. Such surveys are also able to probe galaxy populations irrespective of luminosity, stellar mass, or dust extinction. Additionally, such surveys are sensitive to low-luminosity dwarf systems, which tend to be gas-dominated \citep{2006ApJ...653..240G}. Conversely, such surveys are biased against clusters, the most luminous galaxies, and the `red and dead' galaxy population.

Given the lack of large and deep HI-selected galaxy samples to date (the HIPASS main catalog and its northern extension contain, respectively, 4,315 and 1,002 galaxies; \citet{2004MNRAS.350.1195M,2006MNRAS.371.1855W}), this population, its evolution, and its bias compared to dark matter are poorly understood. The selection of these galaxies is strongly limited in redshift, and targeted observations can only extend to z $\sim$ 0.2 \citep{2008ApJ...685L..13C,2011ApJ...727...40F}, while the Arecibo\footnote{The Arecibo Observatory is operated by SRI International under a cooperative agreement with the National Science Foundation (AST-1100968), and in alliance with Ana G. MŽndez-Universidad Metropolitana, and the Universities Space Research Association.} Legacy Fast ALFA (ALFALFA) survey is limited to z $<$ 0.06. At the same time, this population is poised to become the standard for cosmological measurements based on observations of resolved galaxies as well as intensity mapping. For example, galaxy redshift surveys taking advantage of the 21 cm transition of neutral hydrogen undertaken with instruments like the Square Kilometer Array (SKA) would potentially provide constraints on the dark energy equation of state and its variation with redshift \citep{2010MNRAS.401..743A,2009astro2010S.219M}.

The differences in neutral hydrogen distribution between galaxies in clusters and those in the field are unevenly understood, with proposed solutions spanning from `nature' (i.e., gas-rich galaxies form in low-concentration dark matter halos and/or in underdense environments) to `nurture' (i.e., processes that occur after formation deplete the HI gas from halos, through ram-pressure stripping or galaxy interactions, or enrich HI reservoirs, through cold accretion). The reality is a combination of many processes and initial conditions. Probing the relationship between cold gas mass and other properties known to be anticorrelated with clustering (such as spiral morphology, late type \citep{2002MNRAS.332..827N}, active star formation \citep{2004MNRAS.353..713K}, and blue colors \citep{2005ApJ...630....1Z}) may help to better articulate the influence of environment on galaxy evolution while also constraining the populations to which future large 21 cm line surveys will be sensitive.


Most work directly related to the clustering of gas-rich galaxies came out of the HIPASS survey. \citet{2007ApJ...654..702M} and \citet{2007MNRAS.378..301B} both identified the HIPASS HI-selected sample as the weakest clustering population of galaxies known, but their results regarding the mass dependence of the clustering were in conflict. While the HIPASS team found a statistically insignificant difference between `high' and `low' HI mass galaxies, \citet{2007MNRAS.378..301B} found that high-mass galaxies clustered more strongly. More recently, \citet{2011MNRAS.412L..50P} compare the ALFALFA and HIPASS projected correlation function and angular correlation function, and find that they are similar but that ALFALFA's sensitivity to low-mass galaxies makes that sample more strongly anti-biased relative to dark matter. However, \citet{2011MNRAS.412L..50P}  use only the ALFALFA catalogs published in \citet{2007AJ....133.2569G}, \citet{2008AJ....135..588S}, and \citet{2008AJ....136..713K} ($\sim$ 1,800 galaxies) despite several other ALFALFA catalogs being available at time of publication; these catalogs include the Virgo cluster and Pisces-Perseus foreground void and cover small volumes, so do not comprise a representative sample. \citet{2011MNRAS.412L..50P} are therefore severely limited in their ability to make broader claims about the population.

The excellent sensitivity and large sample size of the $\alpha.40$ sample allows us to probe the clustering characteristics of HI-selected galaxies through the two-point galaxy-galaxy correlation function. 


In the following sections, we describe our dataset (Section \ref{ch6data}) and the methodology used to measure the galaxy-galaxy correlation function (Section \ref{ch6method}). We then estimate the real-space correlation function, both assuming a power law and by direct inversion, and investigate the impact of methodology choices in Section \ref{ch6results}. We compare the ALFALFA clustering results to those found in simulations that have, for the first time, attempted to assign reasonable cold HI gas masses to simulated galaxies, in Section \ref{ch6discussion}, while also discussing the results in context, before concluding in Section \ref{ch6summary}.


\section{Dataset \label{ch6data}}


\subsection{ALFALFA $\alpha.40$ Sample}

The ongoing ALFALFA survey is completing a census of galaxies in the local universe, out to z $\sim$0.06, using the seven-pixel ALFA receiver at the Arecibo Observatory to detect the 21 cm line of neutral hydrogen. Compared to previous blind neutral hydrogen surveys (e.g. HIPASS), ALFALFA's enhanced sensitivity, detection centroiding, volume, and sample size, resulting in a cosmologically representative sample, make it ideally suited for an accurate measurement of the correlation function of gas-rich galaxies.

The sample used here includes the sky coverage of the $\alpha.40$ sample recently presented in \citet{2011AJ....142..170H}, referred to as $\alpha.40$ because it includes the data extracted from coverage of 40$\%$ of ALFALFA's skyprint. The statistical completeness and noise characteristics of the ALFALFA source catalog are well understood and have been discussed extensively elsewhere. Further details may be found both in \citet{2007AJ....133.2087S}, \citet{2010arXiv1008.5107M}, and \citet{2011AJ....142..170H}, which include discussions of the characteristics of the $\alpha.40$ sample and the sensitivity of the ALFALFA survey. In particular, \citet{2011AJ....142..170H} discuss impacts of various volume restrictions on the derivation of the HI mass function. Here we summarize the salient points.

Confidently detected sources in ALFALFA are assigned one of three object codes, where Code 1 refers to a reliable extragalactic detection with a high S/N ($>$ 6.5). For the sample used here, we neglect the other objects, Code 2 and Code 9 sources; Code 9 sources are high velocity clouds (HVCs) of hydrogen gas in the vicinity of the Milky Way and are thus not extragalactic, whereas Code 2 extragalactic sources have lower S/N and are only included in the catalog because they are corroborated by a known optical source at the same position and redshift. Furthermore, ALFALFA's ability to detect extragalactic signal near its redshift limit is degraded due to a strong source of terrestrial radio frequency interference, the FAA radar at the San Juan airport. We therefore include only objects within 15,000 \kms, which results in only a modest loss of source counts.

ALFALFA's sensitivity depends not only on the integrated flux, but also on the 21 cm spectrum's profile width W$_{50}$ (\kms). Because the mass of an HI source is a function of its distance and integrated flux, integrated flux can be thought of as a proxy for mass. The survey therefore is not volume-, flux-, or mass-limited, and the reconstructed selection function must take this complex sensitivity into account. Thus, when the sample is viewed as the distribution of galaxy masses as a function of distance, as in Figure 3 of \citet{2011AJ....142..170H}, it is clear that $\alpha.40$ is sensitive to very low HI mass galaxies nearby but only to significant masses at greater distances

We refer the reader to Figure 1 of \citet{2010arXiv1008.5107M}, which displays the dependence of ALFALFA's sensitivity on both integrated flux and profile width. The HIPASS survey recovered sources with the same dependence on these two parameters. ALFALFA is more sensitive than HIPASS, with a 5$\sigma$ detection limit of 0.72 Jy \kms \, for a source with profile width 200 \kms \, in ALFALFA compared to 5.6 Jy \kms \, for the same source in HIPASS. In \citet{2010arXiv1008.5107M}, we fit a linear relationship between integrated flux and profile width, with a break at 400 \kms, which describes the sensitivity of the survey. We use that same relationship and the selection function derived in that work throughout this paper.



The trimmed sample includes only Code 1 objects from the $\alpha.40$ catalog within 15,000 \kms, for a total of $\sim$10,150 galaxies used in measuring the two-point correlation function.


\subsection{Selection Function}

The distance dependence of the selection function of $\alpha.40$ was determined using the 2DSWML (two-dimensional stepwise maximum likelihood) method, described fully in Appendix B of \citet{2010arXiv1008.5107M}. 2DSWML is related to the SWML (stepwise maximum likelihood) approach, but modified to account for the survey sensitivity's two-dimensional dependence on both integrated flux and profile width. 2DSWML splits the distribution of galaxy masses and profile widths in the $\alpha.40$ sample into logarithmic bins, and then calculates the best-fit HI mass function (analogous to a luminosity function) which maximizes the joint likelihood of detecting all galaxies in the sample. This approach simultaneously measures the selection function for each detected galaxy in the sample. In this work, we will use that selection function $S(D_{i})$ for each galaxy $i$ with a known distance $D_i$ in Mpc.







\begin{figure}[htb]
\begin{center}
\includegraphics[width=4.0in]{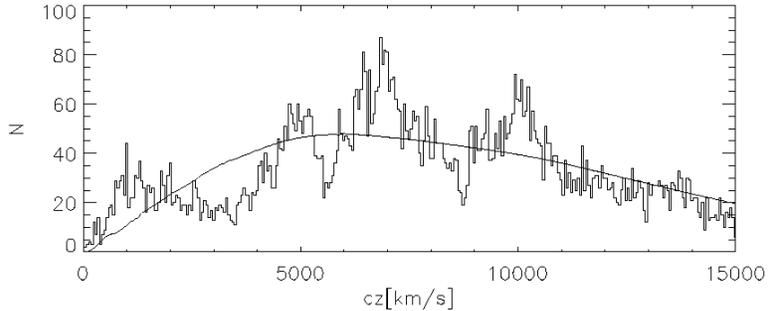}
\caption[Selection function of $\alpha.40$ compared to observations.]{The observed redshift distribution of $\alpha.40$ galaxies (histogram) compared to the expected distribution (solid line) obtained via the survey's selection function. This Figure was previously published as Figure 21 in \citet{2011AJ....142..170H}. \label{ch6fig1}}
\end{center}
\end{figure}

For application to the correlation function, $S_{D_i}$ is calculated for every galaxy in the sample and then $S_{D}$ is smoothed. This smoothed selection function can additionally be combined with an HI mass function to make predictions regarding the number of galaxies of a given mass that are expected to be found in the survey, or used to predict the redshift distribution of the survey under an assumption of homogeneity. Figure \ref{ch6fig1}, previously published in \citet{2011AJ....142..170H}, shows a histogram of the $\alpha.40$ redshift distribution, with peaks and dips representing clusters and voids, respectively, along with an overplotted prediction based on the selection function and a non-clustered Universe.

Disagreements between the prediction and the observations are due both to the existence of large scale structure in the survey volume and to the loss of survey sensitivity at certain velocities due to radio frequency interference. This contamination is quantified as a percentage of survey coverage at a given heliocentric velocity, or spectral weighting, as discussed in \citet{2010arXiv1008.5107M} and earlier publications from the ALFALFA survey. For the purposes described here, the weights map has been translated into the CMB reference frame in order to most accurately model the predicted ALFALFA galaxy distribution (see Section \ref{ch6random}). The selection function is used both for the creation of the random samples for estimation of the correlation function, and for the weighting of pair counts in that estimate.


\section{Method: Estimation of $\xi$(r) and Error Analysis \label{ch6method}}

We measure the correlation function, $\xi(\sigma,\pi)$ in bins of on-sky ($\sigma$) and radial ($\pi$) redshift-space separations, using their observed velocities. Given the redshift extent of $\alpha.40$, we have translated measured galaxy velocities from the heliocentric frame of reference to the CMB frame of reference using \citet{1996ApJ...470...38L}. For two galaxies $i$ and $j$, these separations are:

\begin{equation}
\sigma = \frac{v_i + v_j}{H_0} \times \tan{(\theta/2)}
\label{ch6eqn4}
\end{equation}

and

\begin{equation}
\pi = \frac{|v_i - v_j|}{H_0}
\label{ch6eqn5}
\end{equation}

where $\theta$ is the angular separation of the two galaxies on the sky, $v_i$ and $v_j$ are defined in the CMB reference frame, and H$_0$ is expressed in units of h (H$_0$ = 100h). We adopt the \citet{1983ApJ...267..465D} definitions for $\sigma$ and $\pi$ rather than those used in \citet{1994MNRAS.266...50F}, but note that \citet{1997ApJ...489...37G} found negligible differences for a sample with similar redshift extent and covering a portion of $\alpha.40$'s survey volume. Because we are using a sample of galaxies in the very local universe, we neglect cosmological corrections to the distances. This choice is further supported by our focus on small relative pair distances (always less than 30 h$^{-1}$ Mpc) and our interest in projected quantities where such distance errors, already very small in magnitude, are absorbed in the projection.

Our ultimate goal is to measure $\xi$(r), the real space correlation function, through the observables actually available to us, that is, $\xi(\sigma,\pi)$. In particular, we are interested in modeling the power-law shape of the correlation function up to $\sim$10 h$^{-1}$ Mpc, beyond which point the correlation function is known to diverge from a simple power-law. 

Since $\xi$ measures not simply the probability distribution of galaxy separations in a sample, but the \emph{excess} probability compared to a homogeneously distributed sample, estimators compare the observed galaxy distribution to a random distribution designed to reflect the survey's observational limitations but to exclude the effects of large scale structure. This is straightforwardly accomplished by comparing the number of pairs in ($\sigma$, $\pi$) separation bins from the observed sample to the pair counts from the random sample. In the sections that follow, we will describe this method and the corresponding error analysis in greater detail.


\subsection{Pairwise Estimation\label{ch6pairwise}}

We adopt the Landy-Szalay pairwise estimator \citep{1993ApJ...412...64L} for the correlation function. The Landy-Szalay normalization of data-data (DD), random-random (RR), and data-random (DR) pair counts allows us to construct a random catalog that contains many more objects than the observed data catalog, thereby reducing the introduction of shot noise from the random set. The Landy-Szalay estimator is constructed from these normalized counts: 

\begin{equation}
\hat{\xi}_{LS} = \frac{D_{DD} - 2(D_{DR}) + D_{RR}}{D_{RR}}
\label{ch6eqn10}
\end{equation}

Because $\alpha.40$ is not volume-limited, the pair counts must be weighted so that the measurement is not dominated by galaxies at the peak of the selection function. Following \citet{2007ApJ...654..702M} and \citet{2003MNRAS.346...78H}, we apply a weighting $w_{ij} = w_i \times w_j$ for the contribution of each pair $i,j$ to the Landy-Szalay estimator, given by:

\begin{equation}
w_i = \frac{1.0}{1.0 \: + \: 4\pi \: N_D \: S(r_i) \: J_3(s)}
\label{ch6eqn11}
\end{equation}

where $N_D$ is the number of galaxies in $\alpha.40$, $S(r_i)$ is the selection function measured for $\alpha.40$ at $r_i = cz_{CMB, \: i}/H_0$, and

\begin{equation}
J_3(s) = \int^{s}_{0} s'^2 \: \xi(s') \: ds'
\label{ch6eqn12}
\end{equation}

defined in terms of the redshift space coordinate $s = \sqrt{\sigma^2 + \pi^2}$.

This expression for $J_3$ requires an assumed model for $\xi(s)$, but the final measurement of the correlation function is not sensitive to this assumed input for object weighting; we assume a power-law form:

\begin{equation}
\xi(s) = \left( \frac{s}{s_0} \right )^{-\gamma}
\label{ch6eqn13}
\end{equation}

and we test our robustness by first assuming a fiducial value found for optically-selected samples, $s_0$ = 5.0 h$^{-1}$ Mpc and $\gamma$ = 1.8 and, after that, iterating to the value $s_0$ and $\gamma$ measured for $\alpha.40$. No statistically significant difference is observed through this iterative process, and we therefore proceed as other authors have, using the fiducial optical values reported here in our $J_3$ weighting. Following \citet{1994MNRAS.266...50F, 1994MNRAS.267..927F}, we apply an artificial cutoff with a maximum value of $s$ = 30 h$^{-1}$ Mpc in the expression for $J_3$.


\subsection{Random Samples \label{ch6random}}

\begin{figure}[htb]
\begin{center}
\includegraphics[width=4.0in]{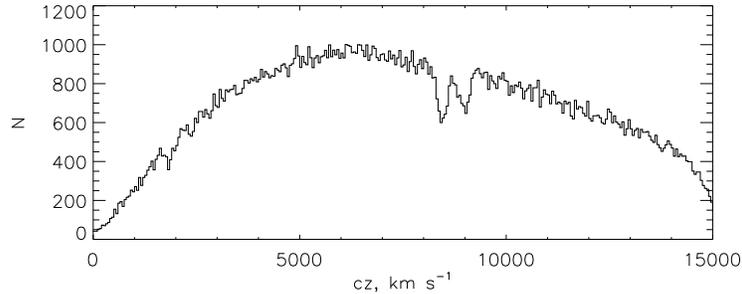}
\caption[Redshift distribution of the random sample.]{The redshift distribution of the constructed random sample. The dips in the distribution at $\sim$ 8,000 \kms \, are due to radio frequency interference at the Arecibo Observatory. When data at these frequencies is flagged as bad (and thus ignored in the processing pipeline), it leads to a reduction in the effective search volume at the corresponding velocities, which translates into a reduction in counts in the random samples. See \citet{2010arXiv1008.5107M} for a plot of the average relative weight as a function of velocity in $\alpha.40$.\label{ch6fig2}}
\end{center}
\end{figure}

We construct random samples that contain 20 times the number of objects in the $\alpha.40$ dataset. These random samples are carefully designed to include survey selection effects while excluding correlations due to large-scale structure. This is accomplished by predicting the distribution of cz$_{CMB}$ from the survey selection and HI mass functions (see Figure \ref{ch6fig1}) and then folding in the loss of volume as a function of velocity due to radio frequency interference, measured from the spectral weights map in \citet{2010arXiv1008.5107M}. Objects in the random set are randomly assigned a sky position within the right ascension and declination boundaries of $\alpha.40$ and are then assigned a redshift from this predicted distribution. The resulting redshift distribution for one example instance of the random sample procedure is shown in Figure \ref{ch6fig2}.

\subsection{Error Analysis}

The correlation function is measured in bins of separation. While the correlation function is expressed as a function of several different coordinates while iterating towards the real-space correlation function $\xi$(r), the bin counts and thus the measured correlation functions are correlated with one another in every such coordinate system. Because structures, such as clusters, will contribute an overabundance of pairs to a set of several bins, the measurement in each bin is not independent of the others. In plots of the correlation function shown here, we display the on-diagonal elements of the covariance matrix (i.e. the standard deviations) as uncertainties on each point. However, in order to work with our measurement to estimate the power-law shape of the correlation function of gas-rich galaxies, we must construct a full covariance matrix and take off-diagonal elements into account.

To construct the covariance matrix $C$, we carry out our pair-counting routine on more than 500 bootstrap resamplings of the data, and a single catalog of random objects is reused in each case. Each of the bootstrap measurements of $\xi$($\sigma$, $\pi$) contains N$_g$ galaxies selected at random from $\alpha.40$, with replacement. From this set of realizations, we construct the covariance matrices for $\xi$(s), $\Xi$($\sigma)$ and $\xi$(r). The covariance between two correlation function bins $b_l$ and $b_m$ is given by:

\begin{equation}
C(l, \: m) = \sum_{i=1}^{N_{realizations}} \frac{(b_{l, \: i} - \bar{b_l})-(b_{m, \: i} - \bar{b_m})}{N_{realizations} - 1}
\label{ch6eqn14}
\end{equation}

The significant off-diagonal elements of the covariance matrix make it difficult to obtain a power-law model fit by minimizing the $\chi^2$ values weighted by the variance. The covariance matrix, however, is not an inescapable quality of the data, but is actually dependent on the basis in which the data are projected. In this case, we have some number of bins N$_b$ representing a set of variables $b$ (bin centers in h$^{-1}$ Mpc), and can choose to work in an orthonormal basis with N$_b$ coordinate axes in which the covariance matrix $C$ is diagonalized. This basis is defined by the principal component eigenvectors of the measurement, and we borrow elements of principal component analysis in order to obtain model parameter fits and uncertainty estimates.

The principal components are linear combinations of the original N$_b$ variables arranged such that the first principal component corresponds to an orthonormal axis through N$_b$-dimensional space that explains the largest proportion of variance in the dataset. These principal component vectors are defined by the eigenvectors of the covariance matrix of the original dataset. 

Following \citet{1994MNRAS.266...50F}, we calculate the principal eigenvectors and construct a diagonalizing matrix, $R$, the columns of which are these eigenvectors, and a new covariance matrix, $\tilde{C}$, projected in the new basis set. Since all of the covariance has been accounted for in the definition of the principal components, $\tilde{C}$ has no off-diagonal elements, and the variance is captured in the on-diagonal elements $\tilde{\sigma}$.

Given $\tilde{C}$ and $R$, a set of models with varying values for $s_0$ and $\gamma$ can be projected into the principal component basis via $\tilde{b}_{model}$ = R$^T$ $b_{model}$, for comparison to the measured $\tilde{b}$. We find the value of each parameter that minimizes the expression

\begin{equation}
\chi^2 = \frac{1}{N_b - 2} \sum^{N_b}_{1} \frac{\tilde{b}_{l} -\tilde{b}_{l,model}}{\tilde{\sigma}_{l}}
\label{ch6eqn15}
\end{equation}

Finally, we construct error ellipses to fully describe the likely parameter space of the power-law model for the correlation function \citep{1992nrfa.book.....P}.

\subsection{Obtaining the Real-Space Correlation Function}

\begin{figure}[htb]
\begin{center}
\includegraphics[width=5.0in]{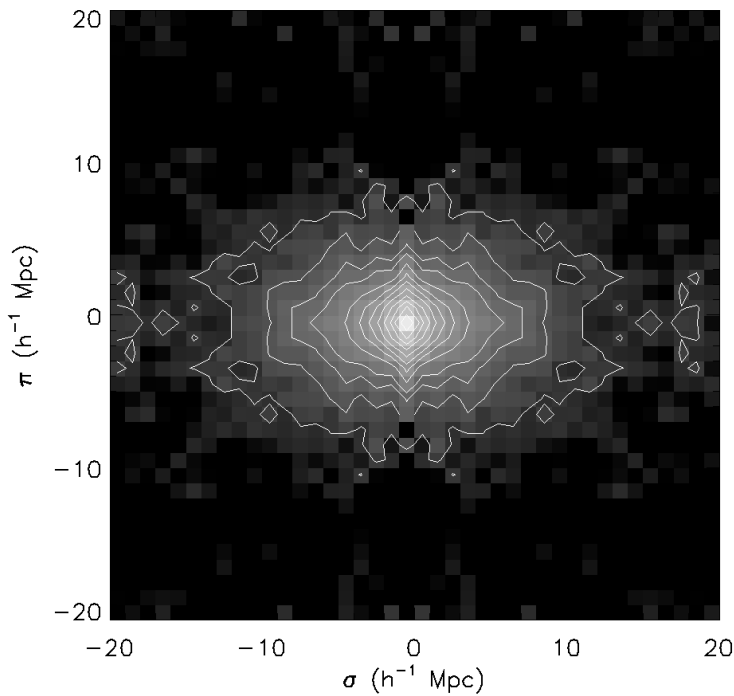}
\caption[Two-dimensional correlation function $\xi(\sigma,\pi)$.]{The two-dimensional correlation function $\xi(\sigma,\pi)$ from $\alpha.40$, measured in h$^{-1}$ Mpc; brighter colors indicate stronger clustering. \label{ch6figtwod}}
\end{center}
\end{figure}

From the three-dimensional galaxy coordinates available to us, we can construct  $\xi(\sigma,\pi)$. This calculation of $\xi(\sigma,\pi)$ is the fundamental measurement upon which the results presented in the rest of this work are based. The resulting image, shown in Figure \ref{ch6figtwod} with contours overplotted, clearly reveals the redshift space distortions that lead to the difficulty in estimating the real-space correlation function. The radial coordinate, $\pi$, appears weakly stretched at small angular separation $\sigma$ because of the Eddington effect in clusters, though because HI-selected galaxies are known to avoid dense cluster regions, this effect is less prominent for $\alpha.40$ than for optically-selected samples. In the other dimension, $\pi$ is flattened on large scales because of the coherent motion of galaxies towards attractors. This `squeezing' effect is determined by the clustering bias (with respect to dark matter) of the sample and by the underlying matter density fluctuations. In a future work, we will explore the shape of $\xi(\sigma,\pi)$ to model the matter density field.

In order to obtain the real-space correlation function $\xi$(r), we must take the intermediate step of projecting $\xi(\sigma,\pi)$ along the $\pi$ axis (in practice, using the discrete bins of size $\Delta \pi$), resulting in what is known as the `projected correlation function' and symbolized as $\Xi(\sigma) / (\sigma)$:

\begin{equation}
\frac{\Xi (\sigma)}{\sigma} = \frac{2}{\sigma} \sum_{0}^{\pi_{max}} \xi (\sigma, \: \pi) \: \Delta \pi
\label{ch6eqn16}
\end{equation}

Following previus work, the maximum value of the integration, $\pi_{max}$, is selected so that the summation is convergent but is kept as low as possible to avoid the introduction of noise from poorly-measured intermediate scales. From the original estimation of $\xi(\sigma,\pi)$, which counted pairs up to distances of $\sim$60 h$^{-1}$ Mpc, we carry the sum along the $\pi$ axis up to a scale of $\pi_{max}$ 29.7 h$^{-1}$ Mpc. Further, we confirm that the resulting correlation function is not strongly sensitive to the chosen value of $\pi_{max}$, but extending the integration to scales that are too large for the sample to sufficiently measure introduces scatter and noise into the correlation function estimate.

$\Xi(\sigma) / (\sigma)$ is closely related to the function in which we are truly interested, $\xi$(r) where r is the real-space distance, via:

\begin{equation}
\frac{\Xi (\sigma)}{\sigma} = \frac{2}{\sigma} \int_{\sigma}^{\infty} \xi(r) \: \frac{r \: dr}{(r^2 \: - \sigma^2)^{1/2}}
\label{ch6eqn17}
\end{equation}

In order to evaluate the real space correlation function, some assumptions must be made about its form. Two options are usually explored in the literature: a power-law form, or a stepwise-function form which makes no assumptions about shape but does assume that the binning used well-represents an underlying smooth correlation function (i.e., the `direct inversion' method). If we assume a power law of the form $\xi$(r) = (r/r$_0$)$^{-\gamma}$, we find:

\begin{equation}
\frac{\Xi (\sigma)}{\sigma} = (\frac{r_0}{\sigma})^{\gamma} \: \frac{\Gamma(1/2) \: \Gamma((\gamma - 1)/2)}{\Gamma(\gamma/2)}
\label{ch6eqn18}
\end{equation}

In Equation \ref{ch6eqn18}, the function $\Gamma$ is the well-known Gamma function. Equation \ref{ch6eqn18} can be recast in terms of fitting parameters:

\begin{equation}
\frac{\Xi (\sigma)}{\sigma} = \left (\frac{r_0}{\sigma} \right )^{\gamma} \: A(\gamma)
\label{ch6eqn19}
\end{equation}

Following \citet{2007ApJ...654..702M}, we rearrange Equation \ref{ch6eqn19}, obtain the best-fit power-law of the form $\xi(\sigma)$ =$a_1 \; \sigma^{a_2}$ using the $\chi^2$ minimization given by Equation \ref{ch6eqn15}, and then relate those parameters to $r_0$ and $\gamma$ which represent the best-fit power law for $\xi$(r).

In the next section, we derive and discuss $\xi$(r) using the mechanisms described in this section.


\section{Results: Clustering in $\alpha.40$ \label{ch6results}}


\subsection{$\Xi(\sigma)/\sigma$ and $\xi$(r) Assuming Power-Law Model \label{ch6fit}}

\begin{figure}[htb]
\begin{center}
\includegraphics[width=3.0in]{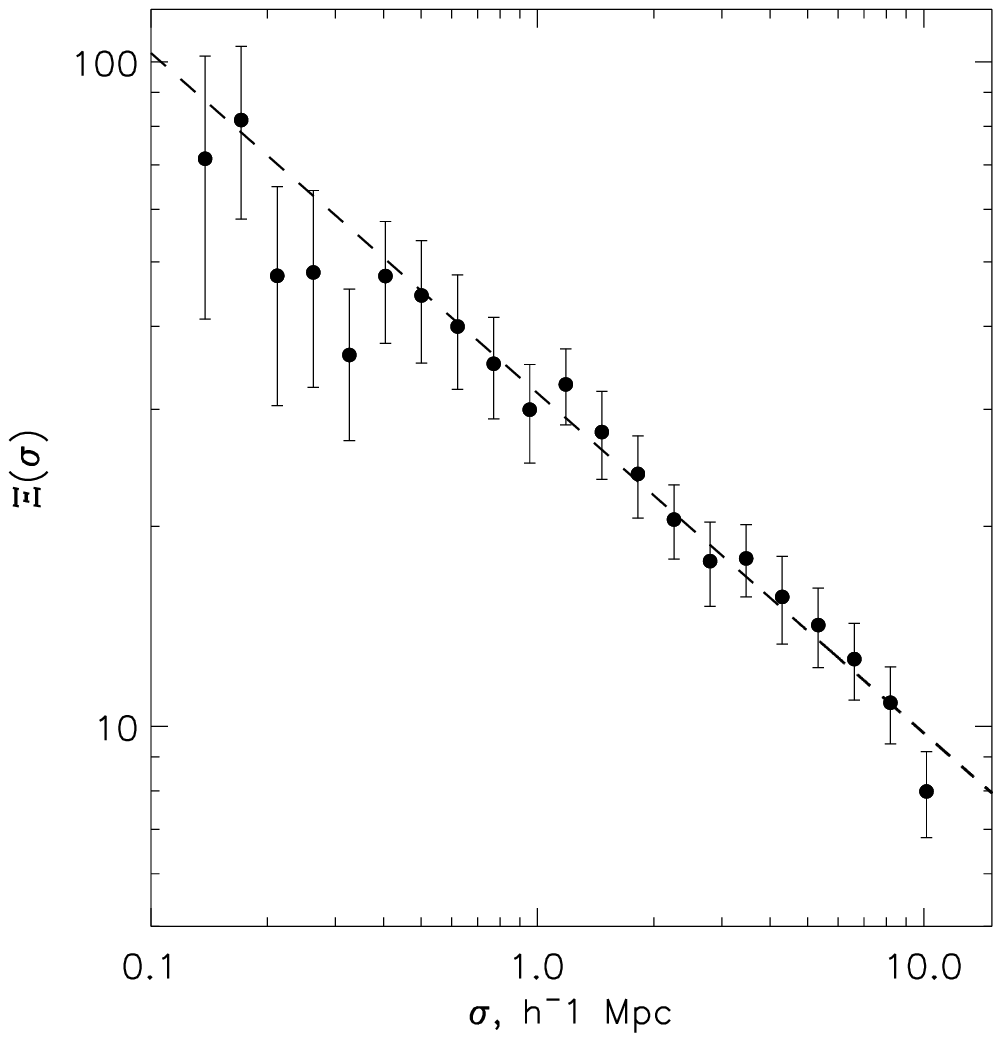}
\caption[Projected correlation function $\Xi(\sigma)$ from $\alpha.40$.]{The projected correlation function $\Xi(\sigma)$ from $\alpha.40$. Error bars reflect the on-diagonal elements of the full covariance matrix. The overplotted dashed line is the fit from the full covariance analysis, with $\gamma$ = 1.51 $\pm$ .09; r$_0$ = 3.3$^{+0.3}_{-0.2}$ (h$^{-1}$ Mpc). \label{ch6fig5}}
\end{center}
\end{figure}

The projected correlation function $\Xi(\sigma)/\sigma$ (recast for the figure and the fitting as $\Xi(\sigma)$) is displayed in Figure \ref{ch6fig5}, along with error bars reflecting the on-diagonal elements of the full covariance matrix. The dashed line is the best-fit model obtained by $\chi^2$ minimization using the full covariance matrix. In Table \ref{ch6tab1}, we list the parameters for the fit and their uncertainties, along with the fits obtained if only the the on-diagonal elements (the standard deviations, $\sigma$) are used to carry out the standard least-squares fit. For comparison, we also include the clustering reported by the HIPASS team (\citet{2007ApJ...654..702M}; note that those authors ignored the off-diagonal elements in their error analysis), the clustering found by \citet{2007MNRAS.378..301B} using the same HIPASS dataset, and the clustering of several optically-selected samples of interest. We also display the \citet{2011MNRAS.412L..50P} results, which used a small, publicly available early subset of the ALFALFA data.

\begin{deluxetable}{ccc}
\tablewidth{0pt}
\tabletypesize{\small}
\tablecaption{Best-Fit Correlation Function Power Law Models\label{ch6tab1}}
\tablehead{ \colhead{Fitting Method} & \colhead{r$_0$ (h$^{-1}$ Mpc)} &  \colhead{$\gamma$} }
\startdata
Full Covariance & 3.3 (+0.3, -0.2) & 1.51 ($\pm$ .09) \\
On-Diagonal Only & 3.2 ($\pm$ 0.1) & 1.48 ($\pm$ .03) \\
Passmoor ALFALFA \tablenotemark{a}& 2.3 ($\pm$ 0.6) & 1.6 ($\pm$ .1) \\
HIPASSa (2007) \tablenotemark{b}& 3.5 ($\pm$ 0.3) & 1.47 ($\pm$ .08) \\
HIPASSb \tablenotemark{c} & 3.3 ($\pm 0.3$) & 1.4 ($\pm$ 0.2) \\
2dFGRS late-type faint\tablenotemark{d} & 3.7 ($\pm 0.8$) & 1.8 ($\pm$ 0.1) \\
SDSS Bright\tablenotemark{e} & 6.2 ($\pm 0.2$) & 1.85 ($\pm$ 0.03) \\
SDSS Faint\tablenotemark{e} & 3.5 ($\pm 0.3$) & 1.92 ($\pm$ 0.05) \\
IRAS All-Sky (real space)\tablenotemark{f} & 3.76 ($\pm$ 0.20) & 1.66 ($\pm$ 0.10) \\
QDOT\tablenotemark{g} & 3.87 ($\pm$ 0.32) & 1.11 ($\pm$ 0.09) \\
Pisces-Perseus early types\tablenotemark{h} & 8.35 ($\pm$ 0.75) & 2.05 ($\pm$ 0.10) \\
Pisces-Perseus late types\tablenotemark{h} & 5.55 ($\pm$ 0.45) & 1.73 ($\pm$ 0.08) \\
\hline
\enddata
\tablenotetext{a}{\citet{2011MNRAS.412L..50P}}
\tablenotetext{b}{\citet{2007ApJ...654..702M}}
\tablenotetext{c}{\citet{2007MNRAS.378..301B}}
\tablenotetext{d}{We include the second-faintest sample due to a warning in \citet{2002MNRAS.332..827N} that the faintest (and smallest) sample provides poorly-constrained fits.}
\tablenotetext{e}{\citet{2005ApJ...630....1Z}}
\tablenotetext{f}{\citet{1994MNRAS.266...50F}}
\tablenotetext{g}{\citet{1994MNRAS.269..742M}}
\tablenotetext{h}{\citet{1997ApJ...489...37G}}
\end{deluxetable}

\begin{figure}[htb]
\begin{center}
\includegraphics[width=4.0in]{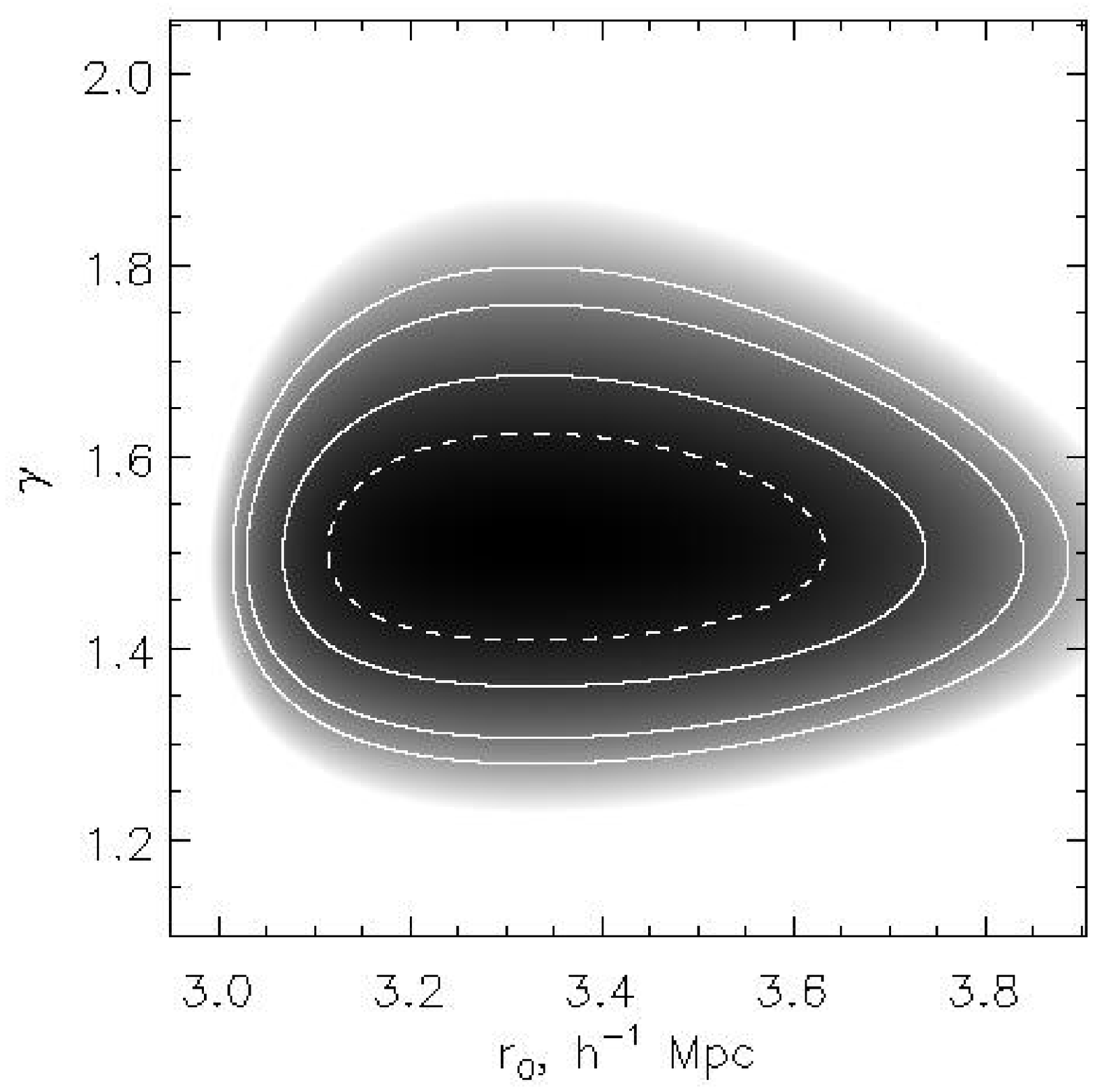}
\caption[$\chi^2$ contours for the fit to $\Xi(\sigma)$.]{$\chi^2$ contours for $\gamma$ and r$_0$ (h$^{-1}$ Mpc). The dashed contour gives the 1$\sigma$ projected uncertainties on $\gamma$ and r$_0$ as single free parameters, and the solid contours give joint 1, 2 and 3 $\sigma$ fits, respectively, to the pair of two free parameters.\label{ch6fig6}}
\end{center}
\end{figure}

Error ellipses are displayed in Figure \ref{ch6fig6}, with the dashed contour giving the 1$\sigma$ single-parameter uncertainties listed in Table \ref{ch6tab1}.

While both the full covariance analysis and the assumption of bin independence give similar results, the larger uncertainties on the full covariance analysis give an indication of the need to be conservative. Parameter uncertainties previously reported in the literature (i.e., \citet{2007ApJ...654..702M}) significantly underestimate their reported statistical uncertainties. Even with the greater sensitivity, larger sample size, and deeper redshift range of $\alpha.40$, the correlation function analysis allows for quite a large range of clustering scenarios.

We confirm the HIPASS result that HI-selected galaxies are among the most weakly clustered known class of galaxies, most comparable to, but still less clustered than, the faint late-type subsampling in 2dFGRS and the IRAS galaxy redshift survey, which was also biased toward starforming galaxies. Similar to the results we will present in Section 6 for $\alpha.40$, the flux-limited sample of IRAS galaxies considered in \citet{1994MNRAS.266...50F} was found to be antibiased relative to cold dark matter on small scales but unbiased on intermediate scales ($\sim$10 h$^{-1}$ Mpc) and positively biased on the largest scales (beyond 10 h$^{-1}$ Mpc). Similarly, the QDOT sample \citep{1994MNRAS.269..742M}, also taken from the IRAS parent catalog but based on a lower flux limit and employing a different sampling strategy, was found to be unbiased with respect to dark matter.

\citet{1997ApJ...489...37G} provides an interesting comparison to our findings, as it was also based on an analysis of 21 cm galaxy profiles observed with the Arecibo Observatory, although that work was based on an optically-selected, magnitude-limited sample rather than a blindly HI-selected sample as in this case. This work also samples a region that partially overlaps with $\alpha.40$. \citet{1997ApJ...489...37G} split their sample by morphological type and determined the variation of clustering strength between early- and late-type (spiral and irregular) galaxies in Zwicky's catalog within the Pisces-Perseus region. They found that the early types were significantly more clustered than the late types, as reflected more generally in Table \ref{ch6tab1}, but their volume-limited sample is significantly more clustered than the HI-selected ALFALFA sample.

Our findings for the clustering of HI-selected galaxies are in agreement with previous results, particularly with our understanding that ALFALFA galaxies tend to be blue, spiral, and late type galaxies which are already known to be weakly clustered \citep{2002MNRAS.332..827N,2004MNRAS.353..713K,2005ApJ...630....1Z}. Apart from the estimates of uncertainties, the clustering of $\alpha.40$ is in agreement with the HIPASS findings but not with \citet{2011MNRAS.412L..50P}, which is not unexpected giving the weaknesses -- in particular extremely small sample size -- of the latter.


\subsection{$\xi$(r) via the Inversion Method and the Shape of the Correlation Function}

$\xi$(r) will only be tidily related to $\Xi(\sigma)/\sigma$ if we assume that the underlying physics of galaxy formation dictates a power-law form for $\xi$(r). In the previous section, we calculated the correlation function of gas-rich galaxies under that assumption, but it is also possible to avoid that assumption and obtain $\xi$(r) by direct inversion of the projected correlation function $\Xi$. The inversion method tests the power-law assumption, though it is a noisy measurement and results in large scatter. As an independent test of the shape of the correlation function, $\xi$(r) determined via this inversion method will be especially useful at scales above $\sim$2.5 h$^{-1}$ Mpc, where the correlation function shows features inconsistent with the power law assumption.

Following \citet{2007ApJ...654..702M}, \citet{2003MNRAS.346...78H} and \citet{1992MNRAS.258..134S}, we take our measurement of $\Xi(\sigma)$ to represent an underlying step function form with values $\Xi_l$ in intervals with centers $\sigma_l$, rearrange Equation \ref{ch6eqn17}, and interpolate between bins to give:

\begin{equation}
\xi(r = \sigma_i)= - \frac{1}{\pi} \sum_{j \ge i} \frac{\Xi(\sigma_{j+1}) - \Xi(\sigma_j)}{\sigma_{j+1} - \sigma_j} \: \times \: \ln \frac{\sigma_{j+1} + \sqrt{\sigma_{j+1}^{2} - \sigma_{i}^{2}}}{\sigma_j + \sqrt{\sigma_{j}^{2} - \sigma_{i}^{2}}}
\label{ch6eqn20}
\end{equation}

The sum in Equation \ref{ch6eqn20} is truncated so that $\sigma_{max} = \pi_{max}$ for the value of $\pi_{max}$ used in Equation \ref{ch6eqn16}.

\begin{figure}[htb]
\begin{center}
\includegraphics[width=3.0in]{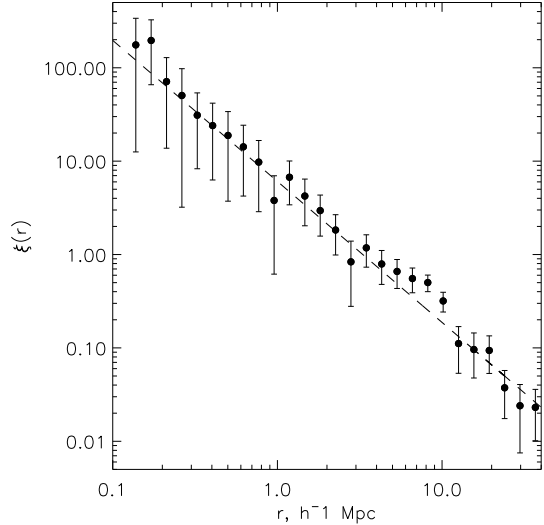}
\caption[$\xi$(r) by the inversion method.]{$\xi$(r) obtained via the inversion method, extended to scales $\sim$ 30 h$^{-1}$ Mpc, with the best-fit power law model for the projected correlation function overplotted as a dashed line to demonstrate agreement.\label{ch6fig7}}
\end{center}
\end{figure}

The projected correlation function and $\xi$(r) obtained by the inversion method are in excellent agreement, as shown in Figure \ref{ch6fig7}, where the points are the inversion $\xi$(r) with on-diagonal uncertainties and the overplotted dashed line is the best-fit power law model for the projected correlation function. It is also clear that the inversion method, as remarked earlier, is more vulnerable to variance. Its use is motivated as a check on the assumed shape of the correlation function. Furthermore, because the power law assumption is known to be useful only on small-to-intermediate scales, the correlation function obtained via the inversion method allows us to extend to large scales, $\sim$ 30 h$^{-1}$ Mpc

By relaxing the power-law assumption, we can further examine divergence from a power-law shape, a well-known phenomenon found in clustering studies of other populations of galaxies. The ALFALFA correlation function shows a `shoulder' at scales of $\sim$ a few h$^{-1}$ Mpc, as observed by, e.g., \citet{1991ApJ...382L...5G, 2003MNRAS.346...78H, 2004ApJ...608...16Z}. Under the assumption of an inflationary, cold dark matter universe and using a halo occupation distribution (HOD) model, \citet{2004ApJ...608...16Z} infer that the well-known shoulder is due to two distinct regimes in which galaxy-galaxy pairs are counted. On large scales, pairs are counted from separate dark matter halos, while on small scales, pairs are counted in the same dark matter halo and are subject to nonlinearity. 

Because the galaxies probed by ALFALFA are gas-rich, and because an HI-selected sample is biased towards gas-dominated low-mass objects that would be classified as LSB (low surface brightness) dwarfs in an optical survey, we expect that the characteristics of the single-halo regime would differ from that observed in the case of an optically-selected sample. For example, tidal interactions and stripping within dense halos, which would decrease the pair counts of HI-selected galaxies, would change the relative contributions of the two regimes.  \citet{2011ApJ...738...22W} find that a resulting power-law correlation function, when the contributions from both the one-halo and two-halo regimes are included, is only found under conditions in a narrow mass and redshift range for the general population of galaxies. Given a different halo occupation distribution model as a function of galaxy properties, the shoulder or break from the power law would be more prominent. The ALFALFA correlation function therefore provides yet another approach from which we can better understand the evolution of HI and the distribution of gas-rich galaxies in the present universe. In a future paper, as the ALFALFA dataset continues to grow, we will present a HOD analysis of the ALFALFA correlation function as an extension of the present work.


\subsection{Systematics and Methodology}


Our estimation of $\xi(\sigma,\pi)$ included two choices upon which our results could be dependent. We selected both the logarithmic binning intervals for pair counting as well as the value $\pi_{max}$ for projtecting $\xi(\sigma,\pi)$ into $\Xi(\sigma)$. Investigations of alternative schemes suggest that our results are not strongly dependent on either the binning or the choice of $\pi_{max}$, though $\pi_{max}$ is specifically selected to lead to a stable solution for $\xi$ without introducing the noise and scatter from scales that are poorly probed by the $\alpha.40$ sample.

We consider whether extreme redshift distortions nearby, for small values of cz$_{CMB}$, could be contaminating our results. To test this possibility, we repeat the measurement, this time excluding galaxies within cz$_{CMB}$ $<$ 2,000 \kms, as well as 3,000 \kms. This reduced the sample size to $\sim$ 9,300 and $\sim$ 8,900, respectively, but we found no difference in the final fitting parameters r$_0$ and $\gamma$. We conclude that there is no advantage to be gained in eliminating nearby galaxies from $\alpha.40$ for the correlation function analysis.

\begin{figure}[htb]
\begin{center}
\includegraphics[width=4.0in]{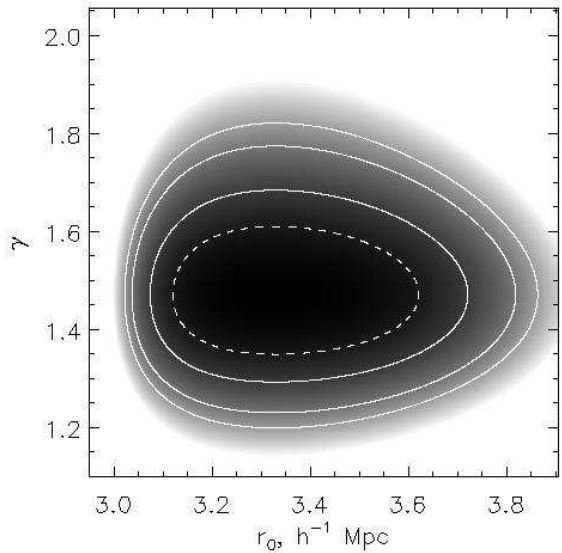}
\caption[$\chi^2$ contours for galaxies with cz$_{CMB}$ $>$ 2000 \kms, using HI parameters for $J_3$.]{$\chi^2$ contours for $\gamma$ and r$_0$ (h$^{-1}$ Mpc) excluding all galaxies with cz$_{CMB}$ $<$ 2,000 \kms. The parameters used in $J_3$ are approximations of the HI-selected r$_0$ (3.4 h$^{-1}$ Mpc) and $\gamma$ (-1.5). The dashed contour gives the 1$\sigma$ projected uncertainties on each parameter, and the solid contours give 1, 2 and 3 $\sigma$ fits, respectively, to the pair of parameters.\label{ch6fig62}}
\end{center}
\end{figure}

The expression for $J_3$ in Equation \ref{ch6eqn12} requires an expression for the shape of the correlation function. In calculating $\xi(\sigma,\pi))$ there is therefore a presumably small dependence on the as-yet unknown parameters $s_0$ and $\gamma$. As briefly mentioned in Section \ref{ch6pairwise}, one possible way to avoid any potential problems is to use the fiducial optical sample parameters $s_0$=5.0 and $\gamma$=-1.8 to calculate the parameters for an HI-selected sample, and then iterate towards a stable solution. In attempts to do this, we find that there is no significant difference between the parameters estimated via these two methods, and confirm that $\xi$ is not dependent on the precise form of $J_3$. Such iteration does not provide any advantage. We demonstrate this in Figure \ref{ch6fig62}, which displays the error contours on the power-law fit parameters for a sample limited to cz$_{CMB}$ $>$ 2,000 \kms, and for $J_3$ using the approximate parameters estimated in Section \ref{ch6fit}. The results are very close to those in Figure \ref{ch6fig6} and the 1$\sigma$ estimated parameters are identical.

\section{Discussion \label{ch6discussion}}

\subsection{Comparison with Mock Catalogs \label{ch6mock}}

The correlation function of gas-rich galaxies has implications for the improvement of galaxy simulations, by providing an observational constraint for the results of simulations. This work will allow a better match between simulations and the observed relationship between gas mass and clustering properties. Simulations are just now progressing to the point where reasonable, realistic cold HI gas masses can be assigned to galaxies. In this section, we will compare the results of the correlation analysis of $\alpha.40$ with presently available cold dark matter simulations.

We are limited in our ability to compare our observations to simulations by what is available publicly. \citet{2010arXiv1008.5107M} took advantage of the \citet{2009ApJ...698.1467O} (hereafter O09) simulation, which assigned cold gas to galaxies from the \citet{2007MNRAS.375....2D} catalog of Millennium Simulation galaxies. In that work, we found that O09's simulation provided a reasonable fit to the observed HI mass function. However, this catalog may not be adequate for comparison with an observed correlation function. In particular, O09 caution that the mass resolution of the simulation prevents them from applying their findings to faint, low surface brightness, or low-mass galaxies. Given the known correlations between galaxy type and clustering, and between HI mass and luminosity, it would be difficult to use this catalog to explore the relationship between current simulations and ALFALFA's observations. Furthermore, O09 did not themselves carry out this analysis.



\citet{2011MNRAS.414.2367K} have provided another option for comparison, using a set of four GALFORM semi-analytical models that treats a range of processes which influence gas reservoirs, including cooling, ram-pressure stripping, mergers, star formation, and supernova feedback. They report results over a range of redshifts, but for this work only their results at z = 0 are of interest. They find that the galaxy-galaxy correlation function of the simulations is consistent with those found for HIPASS, and confirm that their simulation shows gas-rich galaxies as being significantly less clustered than dark matter. Differences between the models and the scales at which those differences are important can be used to highlight potential problems in the assumptions, such as models that overpredict the gas richness of satellites.

\begin{figure}[htb]
\begin{center}
\includegraphics[width=3.0in]{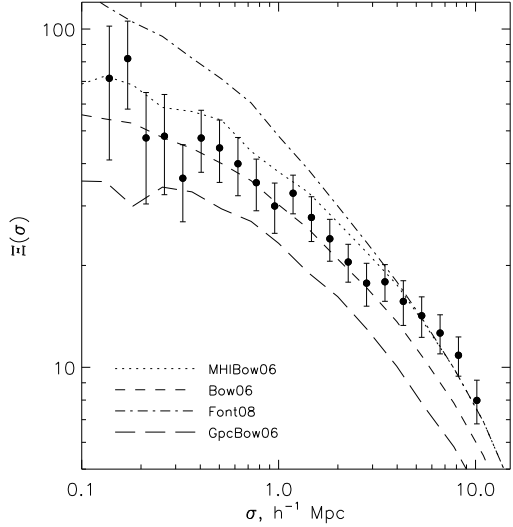}
\caption[Model HI correlation functions compared with $\alpha.40$]{Models for $\xi(\sigma)$ from \citet{2011MNRAS.414.2367K}, compared with $\alpha.40$ (filled points with error bars).\label{ch6figkim}}
\end{center}
\end{figure}

In Figure \ref{ch6figkim}, we compare the models presented in \citet{2011MNRAS.414.2367K} with the observed $\alpha.40$ correlation function for HI-selected galaxies. The models include that of \citet{2006MNRAS.370..645B}, labeled as Bow06; a modified version of the same, labeled MHIBow06; a version that uses a slightly different background cosmology that is in better agreement with the WMAP parameters, labeled GpcBow06; and, finally, the model of \citet{2008MNRAS.389.1619F}, labeled as Font08. In all models, only galaxies with M$_{cold} >$ 10$^{9.5}$ h$^{-2}$ \msun, where M$_{HI}$ = 0.76 M$_{cold}$/(1+.04), are included, which matches the HIPASS galaxy selection but may be more massive than would be ideal for matching $\alpha.40$, which probes very small gas masses \citep{2010arXiv1008.5107M, 2011AJ....142..170H}. The models are described in detail in \citet{2011MNRAS.414.2367K}, and here we only discuss the main differences that may be relevant for a comparison to $\alpha.40$. 

Bow06, MHIBow06, and Font08 all use the Millennium Simulation to track galaxies and halos, while GpcBow06 uses a different method involving merger trees and a large box size. Bow06 and Font08 are able to match optical luminosity functions, but both overpredict the abundance of HI in low-mass galaxies. MHIBow06 was created by adjusting the star formation timescale in Bow06, thereby fixing this excess while maintaining the agreement with optical properties of galaxies. GpcBow06, finally, also has a modified star formation prescription which better fits the HI mass function compared to Bow06.

In Figure \ref{ch6figkim}, it is clear that MHIBow06 and Bow06 fit the observed HI correlation function on small scales, while Font08 drastically overpredicts and GpcBow06 drastically underpredicts the strength of clustering for gas-rich galaxies on those scales. At large scales, both GpcBow06 and Bow06 underpredict the clustering strength, while Font08 and MHIBow06 follow it closely. Although not a perfect match, the MHIBow06 model appears to be most consistent with the clustering of gas-rich galaxies over the full range of accessible scales.

Part of these differences may be due to the mass resolution of the models, given that $\alpha.40$ probes to significantly lower masses than the HIPASS survey for which these models were designed. What is clear, however, is that $\alpha.40$ already provides constraints that can begin to differentiate between successful and unsuccessful models, and the full ALFALFA sample should be able to provide very robust constraints for testing simulations that take HI into account.

\subsection{The Bias Parameter for HI-Selected Galaxies}

The bias between any two classes of objects indicates their relative clustering strength. For cosmological purposes, we are interested in comparing the clustering of types of galaxies with the underlying dark matter halo distribution, in order to understand how well future surveys would probe the true (baryonic + dark) mass distribution. The comparison is achieved through the linear bias parameter at z = 0 $b_0$:

\begin{equation}
\xi_{gal}(r) = b_{0}^2 \: \xi_{DM}(r)
\label{ch6eqn22}
\end{equation}

In general, $b_0$ = $b_0$(r), based on the linear theory prediction that $b_0$ is independent of scale. In real galaxies, however, this is expected to become true only above intermediate scales of $\sim$10 h$^{-1}$ Mpc. If $b_{0} > 1.0$, as is the case for red galaxies which tend to be found in clusters (e.g. \citet{1997ApJ...489...37G, 2002MNRAS.332..827N, 2005ApJ...630....1Z, 2006MNRAS.368...21L, 2008MNRAS.385.1635S}), then the distribution is positively biased with respect to the dark matter. For galaxies like HI-selected populations, $b_{0} < 1.0$ and they are said to be antibiased.

As a proxy for the underlying dark matter distribution at z = 0, we use the correlation function of dark matter halos from the Millennium Simulation, given in \citet{2005Natur.435..629S} as a function over the same scales that we are interested in. The Millennium Simulation, however, used an early WMAP estimated value for the parameter $\sigma_{8}$ of $\sim$ 0.9 that is generally recognized to have been high. We have adjusted our calculations to use the recommended value $\sigma_{8}$ = 0.8 from the Seven-Year WMAP (WMAP7) results \citep{2011ApJS..192...16L}.

\begin{figure}[htb]
\begin{center}
\includegraphics[width=3.in]{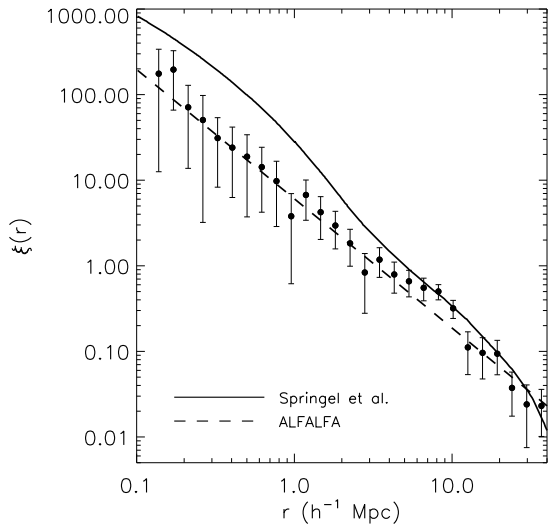}
\caption[Correlation functions for dark matter and HI selected galaxies.]{The real-space correlation function $\xi$(r) for dark matter from the Millennium Simulation (solid line, and adjusted for the WMAP7 value of $\sigma_{8}$) and for HI-selected galaxies from $\alpha.40$ (direct inversion method points with error bars, along with the best-fit power law plotted as dashed line; model fit given in Table \ref{ch6tab1}).\label{ch6fig9}}
\end{center}
\end{figure}

In Figure \ref{ch6fig9} we compare that correlation function to the $\alpha.40$ observation of $\xi$(r) for HI-selected galaxies using the inversion method. The dark matter correlation function deviates strongly from a power law on small scales, an indication of the well-known fact that bias is scale-dependent. Dark matter is, as expected, significantly more strongly clustered than this particular population of galaxies, but the bias becomes less significant at intermediate and large scales.

\begin{figure}[htb]
\begin{center}
\includegraphics[width=4.5in]{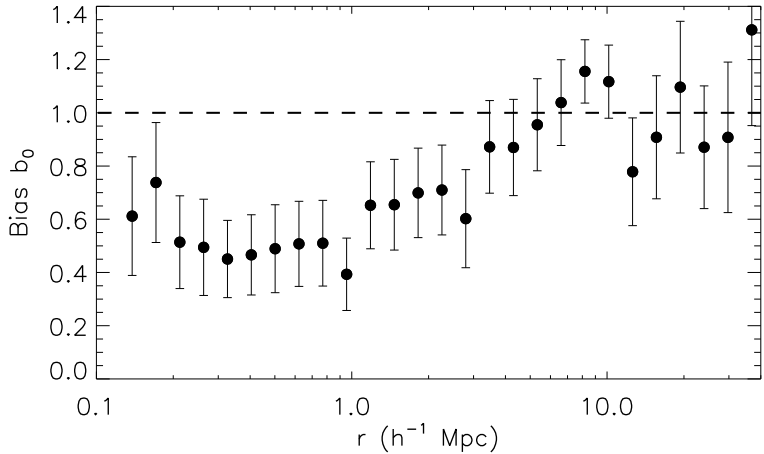}
\caption[The bias parameter b$_0$(r).]{The bias parameter b$_0$(r) as a function of scale.\label{ch6fig10}}
\end{center}
\end{figure}

In Figure \ref{ch6fig10}, we display the bias parameter as a function of scale. The error bars are based only on the $\alpha.40$ uncertainties and they assume that there is no uncertainty in the Millennium Simulation's measurement of the correlation function. This Figure reflects what we already understand about the clustering properties of HI selected galaxies: on small scales, the clustering of gas-rich galaxies is weaker, and on ever-larger scales the distribution of gas-rich galaxies begins to more closely reflect the underlying matter distribution. \citet{2007MNRAS.378..301B} measured the linear bias parameter on large scales for the HIPASS sample using a different technique. Exploring modeled dark matter power spectra for different values of $b_0$, including bias and assuming a concordance cosmology, they identified the most likely bias parameter. They found $b_0$ = 0.7 $\pm$ 0.1, in general agreement with the findings for $\alpha.40$ though we have measured the bias parameter as a function of scale. The preliminary work of \citet{2011MNRAS.412L..50P} is generally consistent with this result, though that work does not capture our finding that the sample becomes unbiased on large scales.

Because of the very limited redshift extent of ALFALFA and the $\alpha.40$ sample, our work cannot comment on the evolution of clustering or bias for HI-selected objects. However, as a robust measurement of these properties at z = 0, we do provide a benchmark observational constraint with which theoretical models will need to agree. The earliest work comparing the bias of HI at z = 0 to the anticipated evolution, by \citet{2007MNRAS.378..301B}, determined b$_0$ $\sim$ 0.68 today and predicted b$_{4}$ would range from $\sim 2 - 4$ by z = 4. Recently, \citet{2010ApJ...718..972M} have used a different, simple bias model, incorporating observational constraints, that relates M$_{HI}$ to M$_{DM}$ to estimate HI masses of Millennium Simulation halos and then investigated the bias of HI with respect to the halo distribution. At z=0, they estimate that the overall linear bias parameter on large scales is $\sim$0.8, slightly higher than our findings at those scales. Their Figure 6 is more comparable to our Figure \ref{ch6fig10}, and shows the same overall rise of the bias with increasing scale found here. Their models also predict that the bias will rise sharply with redshift, with the linear bias parameter reaching b$_{4} \sim$ 2 by z $\sim$ 4.

The $\alpha.40$ observations and \citet{2010ApJ...718..972M} predictions have implications for large-scale 21 cm galaxy surveys and intensity mapping projects with such instruments as the Square Kilometer Array (SKA). If the theoretical results reflect the true evolution of HI gas in the universe, then these projects can expect strong 21 cm signals at a range of redshifts. Perhaps more importantly, the $\alpha.40$ observation of the correlation function at low redshift provides a robust baseline constraint for the development of SKA model predictions, and future simulations will need to match both the HI mass function and the correlation function at z $\sim$ 0. 

Because the HI-selected galaxy bias is likely to be strongly dependent on HI mass, with low-mass objects severely antibiased with respect to the underlying dark matter distribution, high-redshift surveys which are sensitive only to the high-mass end of the HIMF should expect to be mildly antibiased at low redshifts and increasingly positively biased at intermediate to high redshifts. We will explore the mass, color and luminosity dependence of the correlation function as the focus of a future work.


\section{Summary and Conclusions \label{ch6summary}}

We have used the $\sim$10,150 galaxy $\alpha.40$ sample to measure the correlation function of HI-selected galaxies in the local universe. We use bootstrap resampling and a full covariance analysis in order to model the real-space correlation function on scales $<$ 10 h$^{-1}$ Mpc as a power law, $\xi$(r)=(r/r$_0$)$^{-\gamma}$. We find that $\gamma$ = 1.51 $\pm$ 0.09 and that the clustering scale length is r$_0$ = 3.3 + 0.3, -0.2 h$^{-1}$ Mpc. Furthermore, we show using a direct inversion method that the observed $\alpha.40$ real-space correlation function closely follows this power law. The direct inversion method also allows exploration of the divergence from a single power law, seen as a `shoulder' in the correlation function at scales of $\sim$ a few h$^{-1}$ Mpc. Our findings are shown to be robust against the precise form of the weighting used in the pairwise estimation of $\xi(\sigma,\pi)$ and the $\alpha.40$ sample selection criteria. The superior sensitivity of ALFALFA, and high selection function, allows us the include the full survey redshift range (cz = 0 to 15,000 \kms) without the introduction of significant noise in the analysis.

The clustering of HI-selected galaxies is significantly weaker than the clustering of general populations of optically-selected galaxies, and is most closely comparable to samples of faint, late-type, blue and/or starforming galaxies found in optical (and infrared) surveys. Available models of HI in simulated galaxies are in general agreement with our observations, and the $\alpha.40$ measurement of the correlation function is robust enough to begin constraining these models.

Finally, we measure the bias parameter for $\alpha.40$, using the correlation function of dark matter haloes from the Millennium Simulation, and find that the small-scale clustering of HI galaxies is severely antibiased with respect to the underlying dark matter distribution. On large scales the antibiasing becomes only moderate. We suggest that isolating the high-mass galaxies in $\alpha.40$ will show that this population more closely follows the true mass distribution and that an abundance of low-mass galaxies in underdense voids partially explains the strong antibiasing observed. 

The $\alpha.40$ sample provides, for the first time, a robust measurement of the clustering of HI-selected galaxies, which can be used to provide observational constraints for theoretical models. While gas-rich galaxies are, currently, poorly modeled in N-body and semi-analytic simulations of the Universe at z = 0, this situation is likely to change given the results presented here and, especially, the full results when the ALFALFA catalog is complete with a sample of $\sim$ 30,000 objects.

The models of \citet{2011MNRAS.414.2367K}, which reproduce the clustering characteristics of the HIPASS sample, can now be exploited to attempt to understand the clustering revealed by ALFALFA galaxies. Conversely, we may find that these models are not able to reliably reproduce the more complex characteristics of $\alpha.40$, particularly the dependence on galaxy characteristics (e.g., HI mass, color). $\alpha.40$ can therefore contribute to the improvement of these models, working to close the gap between the extremely detailed optical characteristics of simulated galaxies and the poor understanding of where cold gas fits into the picture. To date, such models could only be loosely compared to HI-selected samples, given the lack of a large survey like ALFALFA and robust measurements of cosmological simulations for these samples. Instead, these models typically focus on fitting the luminosities and stellar characteristics of observed galaxies, which are related to gas reservoirs (since gas fuels star formation), but only indirectly.

If simulations can be adjusted now that robust benchmarks exist for the z = 0 characteristics of HI-selected galaxies (e.g., this work, \citet{2010arXiv1008.5107M}, \citet{2011arXiv1106.0710P}, and others), this could constrain the allowed evolutionary tracks that the distribution of gas reservoirs may have followed. Further, the clustering of HI-selected galaxies, in particular high-mass galaxies, can be applied to make predictions of the strength of the signal that will be obtained with future intensity mapping projects, which will not resolve individual galaxies but will measure the bulk HI on $\sim$10 Mpc scales. The $\alpha.40$ measurement of the HI-selected galaxy bias indicates that, at low redshift, the selection of high-HI mass galaxies over large scales ensures a sample that adequately probes the underlying dark matter distribution.

These findings, and the potential for more robust understanding of the role of gas in galaxy evolution, motivate further work in this area. We will further explore clustering properties of gas-rich galaxies as the focus of future work; Papastergis et al. 2011 (in preparation) are analyzing the dependence of $\xi(r)$ on such properties as galaxy color, gas fraction, luminosity, and HI mass. ALFALFA has a distinct advantage in exploring this dependence, given its high sensitivity across 5 orders of magnitude in HI mass, its blind ability to detect both low surface brightness and extremely large, bright spirals, its coverage of a cosmologically representative volume, and its overall sample size.


\acknowledgements
The authors would like to acknowledge the work of the entire ALFALFA collaboration team in observing, flagging, and extracting the catalog of galaxies used in this work. We also thank Han-Seek Kim for kindly providing the correlation function models from GALFORM, presented in \citet{2011MNRAS.414.2367K}, and we acknowledge the efforts of the GALFORM team in developing the code that led to those model results.

This work was supported by NSF grants AST-0607007 and AST-1107390, and by grants from the National Defense Science and Engineering Graduate (NDSEG) fellowship and from the Brinson Foundation. AMM was partially supported by an appointment to the NASA Postdoctoral Program at the LaRC, administered by Oak Ridge Associated Universities through a contract with NASA.

\bibliographystyle{apj}
\bibliography{myreferences}

\end{document}